\documentclass[twocolumn,amsmath,amssymb,floatfix]{revtex4}

\usepackage{bm}
\usepackage{amsmath}

\ifx\epsfannounce\undefined \def\epsfannounce{\immediate\write16}\fi
 \epsfannounce{This is `epsf.tex' v2.7k <10 July 1997>}%
\newread\epsffilein    % file to \read
\newif\ifepsfatend     % need to scan to LAST %%BoundingBox comment?
\newif\ifepsfbbfound   % success?
\newif\ifepsfdraft     % use draft mode?
\newif\ifepsffileok    % continue looking for the bounding box?
\newif\ifepsfframe     % frame the bounding box?
\newif\ifepsfshow      % show PostScript file, or just bounding box?
\epsfshowtrue          % default is to display PostScript file
\newif\ifepsfshowfilename % show the file name if \epsfshowfalse specified?
\newif\ifepsfverbose   % report what you're making?
\newdimen\epsfframemargin % margin between box and frame
\newdimen\epsfframethickness % thickness of frame rules
\newdimen\epsfrsize    % vertical size before scaling
\newdimen\epsftmp      % register for arithmetic manipulation
\newdimen\epsftsize    % horizontal size before scaling
\newdimen\epsfxsize    % horizontal size after scaling
\newdimen\epsfysize    % vertical size after scaling
\newdimen\pspoints     % conversion factor
\def\epsfbox#1{\global\def\epsfllx{72}\global\def\epsflly{72}%
   \global\def\epsfurx{540}\global\def\epsfury{720}%
   \def\lbracket{[}\def\testit{#1}\ifx\testit\lbracket
   \let\next=\epsfgetlitbb\else\let\next=\epsfnormal\fi\next{#1}}%
\def\epsfgetlitbb#1#2 #3 #4 #5]#6{%
   \epsfgrab #2 #3 #4 #5 .\\%
   \epsfsetsize
   \epsfstatus{#6}%
   \epsfsetgraph{#6}%
}%
\def\epsfnormal#1{%
    \epsfgetbb{#1}%
    \epsfsetgraph{#1}%
}%
\newhelp\epsfnoopenhelp{The PostScript image file must be findable by
TeX, i.e., somewhere in the TEXINPUTS (or equivalent) path.}%
\def\epsfgetbb#1{%
    \openin\epsffilein=#1
    \ifeof\epsffilein
        \errhelp = \epsfnoopenhelp
        \errmessage{Could not open file #1, ignoring it}%
    \else                       %process the file
        {%                      %start a group to contain catcode changes
            \chardef\other=12
            \def\do##1{\catcode`##1=\other}%
            \dospecials
            \catcode`\ =10
            \epsffileoktrue         %true while we are looping
            \epsfatendfalse     %[02-Jul-1996]: add forgotten initialization
            \loop               %reading lines from the EPS file
                \read\epsffilein to \epsffileline
                \ifeof\epsffilein %then no more input
                \epsffileokfalse %so set completion flag
            \else                %otherwise process one line
                \expandafter\epsfaux\epsffileline:. \\%
            \fi
            \ifepsffileok
            \repeat
            \ifepsfbbfound
            \else
                \ifepsfverbose
                    \immediate\write16{No BoundingBox comment found in %
                                    file #1; using defaults}%
                \fi
            \fi
        }%                      %end catcode changes
        \closein\epsffilein
    \fi                         %end of file processing
    \epsfsetsize                %compute size parameters
    \epsfstatus{#1}%
}%
\def\epsfclipoff{\def\epsfclipstring{\ifepsfdraft\space clip\fi}}%
\epsfclipoff % default for dvips is OFF
\def\epsfspecial#1{%
     \epsftmp=10\epsfxsize
     \divide\epsftmp\pspoints
     \ifnum\epsfrsize=0\relax
       \includegraphics{\ifepsfdraft}%
     \else
       \epsfrsize=10\epsfysize
       \divide\epsfrsize\pspoints
       \includegraphics{\ifepsfdraft}%
     \fi
}%
\def\epsfframe#1%
{%
  \leavevmode                   % so we can put this inside
  \setbox0 = \hbox{#1}%
  \dimen0 = \wd0                                % natural width of argument
  \advance \dimen0 by 2\epsfframemargin         % plus width of 2 margins
  \advance \dimen0 by 2\epsfframethickness      % plus width of 2 rule lines
  \vbox
  {%
    \hrule height \epsfframethickness depth 0pt
    \hbox to \dimen0
    {%
      \hss
      \vrule width \epsfframethickness
      \kern \epsfframemargin
      \vbox {\kern \epsfframemargin \box0 \kern \epsfframemargin }%
      \kern \epsfframemargin
      \vrule width \epsfframethickness
      \hss
    }% end hbox
    \hrule height 0pt depth \epsfframethickness
  }% end vbox
}%
\def\epsfsetgraph#1%
{%
   \leavevmode
   \hbox{% so we can put this in \begin{center}...\end{center}
     \ifepsfframe\expandafter\epsfframe\fi
     {\vbox to\epsfysize
     {%
        \ifepsfshow
            \vfil
            \hbox to \epsfxsize{\epsfspecial{#1}\hfil}%
        \else
            \vfil
            \hbox to\epsfxsize{%
               \hss
               \ifepsfshowfilename
               {%
                  \epsfframemargin=3pt % local change of margin
                  \epsfframe{{\tt #1}}%
               }%
               \fi
               \hss
            }%
            \vfil
        \fi
     }%
   }}%
   \global\epsfxsize=0pt
   \global\epsfysize=0pt
}%
\def\epsfsetsize
{%
   \epsfrsize=\epsfury\pspoints
   \advance\epsfrsize by-\epsflly\pspoints
   \epsftsize=\epsfurx\pspoints
   \advance\epsftsize by-\epsfllx\pspoints
   \epsfxsize=\epsfsize{\epsftsize}{\epsfrsize}%
   \ifnum \epsfxsize=0
      \ifnum \epsfysize=0
        \epsfxsize=\epsftsize
        \epsfysize=\epsfrsize
        \epsfrsize=0pt
      \else
        \epsftmp=\epsftsize \divide\epsftmp\epsfrsize
        \epsfxsize=\epsfysize \multiply\epsfxsize\epsftmp
        \multiply\epsftmp\epsfrsize \advance\epsftsize-\epsftmp
        \epsftmp=\epsfysize
        \loop \advance\epsftsize\epsftsize \divide\epsftmp 2
        \ifnum \epsftmp>0
           \ifnum \epsftsize<\epsfrsize
           \else
              \advance\epsftsize-\epsfrsize \advance\epsfxsize\epsftmp
           \fi
        \repeat
        \epsfrsize=0pt
      \fi
   \else
     \ifnum \epsfysize=0
       \epsftmp=\epsfrsize \divide\epsftmp\epsftsize
       \epsfysize=\epsfxsize \multiply\epsfysize\epsftmp
       \multiply\epsftmp\epsftsize \advance\epsfrsize-\epsftmp
       \epsftmp=\epsfxsize
       \loop \advance\epsfrsize\epsfrsize \divide\epsftmp 2
       \ifnum \epsftmp>0
          \ifnum \epsfrsize<\epsftsize
          \else
             \advance\epsfrsize-\epsftsize \advance\epsfysize\epsftmp
          \fi
       \repeat
       \epsfrsize=0pt
     \else
       \epsfrsize=\epsfysize
     \fi
   \fi
}%
\def\epsfstatus#1{% arg = filename
   \ifepsfverbose
     \immediate\write16{#1: BoundingBox:
                  llx = \epsfllx\space lly = \epsflly\space
                  urx = \epsfurx\space ury = \epsfury\space}%
     \immediate\write16{#1: scaled width = \the\epsfxsize\space
                  scaled height = \the\epsfysize}%
   \fi
}%
{\catcode`\%=12 \global\let\epsfpercent=%\global\def\epsfbblit{%BoundingBox}}%
\global\def\epsfatend{(atend)}%
\long\def\epsfaux#1#2:#3\\%
{%
   \def\testit{#2}%             % save second character up to just before colon
   \ifx#1\epsfpercent           % then first char is percent (quick test)
       \ifx\testit\epsfbblit    % then (slow test) we have %%BoundingBox
            \epsfgrab #3 . . . \\%
            \ifx\epsfllx\epsfatend % then ignore %%BoundingBox: (atend)
                \global\epsfatendtrue
            \else               % else found %%BoundingBox: llx lly urx ury
                \ifepsfatend    % then keep parsing ALL %%BoundingBox lines
                \else           % else stop after first one parsed
                    \epsffileokfalse
                \fi
                \global\epsfbbfoundtrue
            \fi
       \fi
   \fi
}%
\def\epsfempty{}%
\def\epsfgrab #1 #2 #3 #4 #5\\{%
   \global\def\epsfllx{#1}\ifx\epsfllx\epsfempty
      \epsfgrab #2 #3 #4 #5 .\\\else
   \global\def\epsflly{#2}%
   \global\def\epsfurx{#3}\global\def\epsfury{#4}\fi
}%
\def\epsfsize#1#2{\epsfxsize}%
\def\lsim{\mathrel{\mathstrut\smash{\ooalign{\raise2.5pt\hbox{$<$}\cr\lower2.5pt\hbox{$\sim$}}}}}
\def\gsim{\mathrel{\mathstrut\smash{\ooalign{\raise2.5pt\hbox{$>$}\cr\lower2.5pt\hbox{$\sim$}}}}}

\def\myputfigure#1#2#3#4#5%
{\vskip#5pt\makebox[0pt]{\hskip#2in
\includegraphics[width=#3\textwidth]{#1}}\vskip#4pt\hfill}

\def\beb{\begin{thebibliography}{999}}
\def\eeb{\end{thebibliography}}

\def\be{\begin{equation}}
\def\ee{\end{equation}}
\def\bea{\begin{eqnarray}}
\def\eea{\end{eqnarray}}

\def\cnfw{c_{\rm nfw}}

\begin{document} 

\title{Constraining the Evolution of Dark Energy with a Combination of \\ Galaxy Cluster Observables}

\author{Sheng Wang$^{1,2}$, Justin Khoury$^{3}$, Zolt\'{a}n Haiman$^{4}$ and Morgan May$^{1}$}

\affiliation{$^1$Brookhaven National Laboratory, Upton, NY 11973--5000, USA \\
$^2$Department of Physics, Columbia University, New York, NY 10027, USA \\
$^3$Institute for Strings, Cosmology and Astroparticle Physics, Columbia University, New York, NY 10027, USA \\
$^4$Department of Astronomy, Columbia University, New York, NY 10027, USA}

\begin{abstract}
We show that the abundance and redshift distribution ($dN/dz$) of
galaxy clusters in future high--yield cluster surveys, combined with
the spatial power spectrum ($P_c(k)$) of the same clusters, can place
significant constraints on the evolution of the dark energy equation
of state, $w=w(a)$. We evaluate the expected errors on $w_a=-dw/da$
and other cosmological parameters using a Fisher matrix approach, and
simultaneously including cluster structure evolution parameters in our
analysis. We study three different types of forthcoming surveys that
will identify clusters based on their X--ray emission (such as DUO,
the Dark Universe Observatory), their Sunyaev--Zel'dovich (SZ)
decrement (such as SPT, the South Pole Telescope), or their weak
lensing (WL) shear (such as LSST, the Large Synoptic Survey Telescope).  We
find that combining the cluster abundance and power spectrum
significantly enhances constraints from either method alone.  
We show that the weak-lensing survey can deliver a constraint as
tight as $\Delta w_a \sim 0.1$ on the evolution of the dark energy
equation of state, and that the X--ray and SZ surveys each yield
$\Delta w_a \sim 0.4$ separately, or $\Delta w_a \sim 0.2$ when
these two surveys are combined. 
For the X--ray and SZ surveys, constraints on dark energy parameters
are improved by a factor of two by combining the cluster data with
cosmic microwave background (CMB) anisotropy measurements by Planck,
but degrade by a factor of two if the survey is required to solve
simultaneously for cosmological and cluster structure evolution
parameters. The constraint on $w_a$ from the weak lensing survey
is improved by $\sim 25$\% with the addition of Planck data.
\end{abstract}

\maketitle

\section{Introduction}

It has long been realized that clusters of galaxies provide a useful
probe of fundamental cosmological parameters.  The formation of the
massive dark matter potential wells is dictated by simple
gravitational physics, and the abundance and redshift distribution of
clusters ($dN/dz$) should be determined by the geometry of the
universe and the power spectrum of initial density fluctuations.
Early studies of nearby clusters used this relation to constrain the
amplitude $\sigma_8$ of the power spectrum
({\it e.g.},~\cite{white93,viana96}).  Subsequent works
({\it e.g.},~\cite{bahcall98,blanchard98,viana99}) have shown that the
redshift--evolution of the observed cluster abundance places useful
constrains on the matter density parameter $\Omega_m$.

The next generation of surveys, utilizing the Sunyaev--Zel'dovich effect
(SZE), X--ray flux, or weak lensing signatures to identify galaxy clusters, will be able to deliver large
catalogs, containing many thousands of clusters, with complementary
selection criteria.  Such forthcoming datasets have rekindled a strong
theoretical and experimental interest in galaxy clusters. In particular, Wang \&
Steinhardt~\cite{wang98} first argued that the cluster abundance can be used
to probe the properties of dark energy, and Haiman, Mohr \& Holder~\cite{haiman01} 
showed that a survey with several thousand clusters can yield
precise statistical constraints on both its density $(\Omega_{\rm DE}$) and
its equation of state ($w\equiv P/\rho$). Several subsequent recent
works have focused on various aspects of extracting cosmological
parameters from high--yield, future surveys, such as the statistical
constraints available on curvature $\Omega_k$~\cite{holder01}; 
assessing the impact of sample variance~\cite{hukrav03}
and other uncertainties~\cite{levine02} on parameter estimates;
and controlling such uncertainties by utilizing information from the
shape of the cluster mass function $dN/dM$~\cite{hu03}. Closest to the
subject of the present paper, Weller {\it et al.}~\cite{weller02} and Weller \&
Battye~\cite{weller03} considered constraints on the time evolution of the dark
energy equation of state in forthcoming SZE cluster surveys.

Recent studies have elucidated the additional cosmological information
available from the spatial distribution of galaxy clusters through a
measurement of their three--dimensional power spectrum $P_c(k)$~\cite{huha03}
(see~\cite{matsubara02} for a more general treatment
of extracting cosmological information from redshift surveys).  The
power spectrum contains cosmological information from the intrinsic
shape of the transfer function~\cite{refregier02} and from baryon
features~\cite{blake03,seo03,linder03}.
The existing sample of $\sim 400$ nearby clusters in the REFLEX
survey has already been used to derive their power spectrum;
combined with the number counts, this has yielded constraints on
$\sigma_8$ and $\Omega_m$ that are consistent with other recent
determinations~\cite{schuecker03}.

Most importantly, the cluster abundance and power spectrum can provide
two independent powerful probes of cosmological parameters from a
single dataset. The dependence of $dN/dz$ and $P_c(k)$ on the
cosmological parameters are different. Combining the two pieces of
information can therefore break degeneracies present in either method
alone, and yield tighter statistical constraints. Furthermore, this
can be used to significantly reduce systematic errors arising from
the mass--observable relation, making cluster surveys
``self--calibrating''~\cite{majumdar04}.  This self--calibration is
especially strong when the abundance and power spectrum information
is combined with even a modest follow--up mass calibration
program~\cite{majumdar03}.

In this paper, we return to the question of constraining the
time--evolution of the dark energy equation of state.  Specifically,
we ask the question: {\it can improved constraints be obtained on the
time--evolution of $w=w(z)$ when the cluster counts and power spectrum
are combined?}  We quantify the statistical constraints
expected to be available from future samples of $\gsim 10,000$ galaxy
clusters.

We study constraints from three different types of forthcoming cluster surveys.
The proposed DUO ({\it Dark Universe Observatory})~\cite{duo} X--ray survey will
be performed by an Earth--orbiting satellite consisting of seven telescopes that take
a wide survey of the sky in soft X--ray bands.
The SPT--like ({\it South Pole Telescope})~\cite{spt} SZE survey will
be performed by an 8--meter precision submillimeter--wave telescope detecting distant
galaxy clusters by their Sunyaev--Zel'dovich decrement.
The LSST--like  ({\it Large Synoptic Survey Telescope})~\cite{lsst} survey
will be performed by a ground--based telescope detecting clusters by their
weak lensing shear signature on background galaxies.

The most important differences between the present paper and earlier works that have addressed the time--evolution of the equation of
state~\cite{weller02,weller03} are that here (i) we
simultaneously include the abundance and power spectrum in our
analysis; (ii) in addition to the cosmological parameters, we simultaneously include several
parameters that describe cluster structure and evolution; and (iii) we
study three different types of forthcoming surveys. We also chose a different
fiducial model for our analysis (one close to the standard ``concordance'' cosmology). Our calculational
method, based on Fisher matrices, is, on the other hand, only a simple approximation to the 
Monte--Carlo likelihood analysis performed in~\cite{weller02,weller03}.

This paper is organized as follows.  In \S\ref{sec:method}, we
describe our basic calculational methods.  In \S\ref{sec:results}, we
present our results for different future cluster surveys.
In \S\ref{sec:discussion}, we critically discuss our results,
including their uncertainties, and summarize the implications of this work.

\section{Calculational Method}
\label{sec:method}

\subsection{Simulating Cluster Surveys}
\label{subsec:sim}

We follow the standard approach, and identify galaxy clusters with
dark matter halos. The differential comoving number density of
clusters is given by Jenkins {\it et al.}~\cite{jenkins01}
\bea
\nonumber
\frac{d n}{d M}(z,M)& = & 0.301 \frac{\rho_m}{M} \frac{d \ln 
\sigma^{-1}(M,z)}{d M} \\
&& \times \exp [-|\ln \sigma^{-1}(M,z) + 0.64 |^{3.82}]\,, 
\label{eq:dndz}
\eea
where $\sigma^2 (M,z)$ is the variance of the linear density field at
redshift $z$, smoothed with a spherical top--hat filter which would
enclose mass $M$ at the mean present--day matter density
$\rho_m$~\footnote{Note that we use the fitting formula describing the
unsmoothed mass function in the simulations, given in Eq.~(B3) in
Jenkins {\it et al.}~\cite{jenkins01}, which is more appropriate to
galaxy clusters than the smoothed mass function~\cite{hukrav03}.}.
The Jenkins {\it et al.} mass function was derived from numerical
simulations, and its self--similar form is demonstrated to be accurate
to within $\sim 15\%$ in three widely separated cosmologies (although
see \cite{battye03} who find a more significant cosmology--dependence
of the mass function). Jenkins {\it et al.}  identify simulated
clusters using $M_{180}$, the mass enclosed within a spherical
overdensity of $180$ with respect to the {\it mean} matter
density. However, it is customary to define the relation between
X--ray or SZE flux and halo mass $M_{200}$, defined as the cluster
mass enclosed within a sphere with mean interior overdensity of 200
relative to the {\it critical} density.  To combine this relation with
the mass function in Eq.~(\ref{eq:dndz}), we convert $M_{200}$ to
$M_{180}$ assuming that the halo density profile is described by the
NFW model with a concentration parameter of $\cnfw =
5$~\cite{navarro97}.

The spatial distribution of clusters is assumed to follow the
spatial distribution of the dark matter halos and is specified by the
cluster power spectrum $P_c(k)$.  We follow Hu \& Haiman~\cite{huha03}
and obtain $P_c(k)$ from the underlying mass power spectrum, $P(k)$,
modified by redshift--space distortions~\cite{kaiser87}
\be 
P_c(k_{\perp},k_{\parallel}) = \left[1 + \beta 
\left(\frac{k_{\parallel}}{k}\right)^2\right]^2 ~ b^2 ~ P(k)\,, 
\label{eq:pofk}
\ee
\be 
k^2 = k_{\perp}^2 + k_{\parallel}^2\,, 
\ee
where $k_\perp$ and $k_\parallel$ are the wavenumbers of the
sinusoidal fluctuation modes transverse and parallel to the line of
sight, respectively. The redshift--distortion parameter $\beta$ is
defined by~\cite{kaiser87}
\be 
\beta = \frac{1}{b} \frac{d \ln D_{\rm grow}}{d \ln a}\,,
\label{eq:beta}
\ee
where $D_{\rm grow}$ is the linear growth rate, and $a$ is the expansion
factor normalized to unity today. The parameter $b$ in Eqs.~(\ref{eq:pofk})
and~(\ref{eq:beta}) represents the linear bias averaged over all halos at
redshift $z$:
\be b(z) = \int_{M_{\rm min}(z)}^{\infty} \frac{d n(M,z)}{d M} b(M)\,dM 
\left[ 
\int_{M_{\rm min}(z)}^{\infty} \frac{d n}{d M} \,dM
\right]^{-1}
\label{eq:biasz}
\,, \ee
where $M_{\rm min}(z)$ is the mass of the smallest detectable cluster,
which depends on the type of survey as discussed in
\S\ref{sec:method}.D. The bias parameter of halos of a fixed mass $M$,
$b(M)$, is assumed to be scale independent and given by
\be b(M) = 1 + \frac{a \delta_c^2 / \sigma^2 - 1}{\delta_c} +
\frac{2 p/ \delta_c}{1 + (a \delta_c^2 / \sigma^2)^p}\,, \ee
with $a = 0.75$ and $p = 0.3$ providing the best fits to the
clustering measured in cosmological simulations~\cite{sheth99}.
Finally, $\delta_c$ represents the threshold linear overdensity
corresponding to spherical collapse, whose value is $\delta_c = 1.686$
in an Einstein--de Sitter universe. We keep it fixed throughout the
calculation, as it was shown to be only weakly dependent on cosmology
and redshift in other models~\cite{wang98}.

\subsection{Fisher Matrix Technique}

The Fisher matrix formalism allows a forecasting of the ability of a
given survey to constrain cosmological parameters~\cite{tegmark97}.
It gives a lower bound to the statistical uncertainty of each model
parameter that is to be fit by future data. The well--known advantages
of the Fisher matrix technique are that (i) it allows a quick estimate
of errors in a multi--dimensional parameter space, since the likelihood
functions do
not have to be evaluated at each point of the multi--dimensional grid, and (ii) 
constraints from independent data sets or methods can be easily combined by
simply summing the individual Fisher matrices.

The Fisher matrix is defined as
\be 
F_{ij} = \left<
\frac{\partial^2 \mathcal{L}}{\partial p_i \partial p_j}
\right>,
\ee
where $\mathcal{L} = - \ln L$ is the log--likelihood function, and where $p_i$'s
are the various model parameters which, in our case, include both cosmological
parameters and parameters describing cluster structure and evolution.
The inverse $(F^{-1})_{ij}$ gives the best attainable covariance matrix,
regardless of any specific method used to estimate the parameters
from the data~\cite{tegmark97}. In particular, the best statistical
uncertainty attainable on any individual parameter $p_i$, after
marginalization over all other parameters, is $(F^{-1})_{ii}^{1/2}$.

We construct the Fisher matrix for the redshift distribution of the number density
of galaxy clusters as~\cite{holder01}
\be 
F_{\mu \nu}^{{\rm counts}} = \sum_{i} \frac{\partial N_i}{\partial p_{\mu}} 
\frac{\partial N_i}{\partial p_{\nu}} \frac{1}{N_i}\,,
\label{eq:countfm} 
\ee
where
\be 
N_i = \Delta \Omega \Delta z \frac{d^2 V}{d z d \Omega}(z_i) 
\int_{M_{\rm min}(z_i)}^{\infty} \frac{d n(M,z_i)}{d M} \,dM
\label{eq:nofz} 
\ee
is the number of clusters above the detection threshold in each
redshift bin $i$ centered at $z_i$. In Eq.~(\ref{eq:nofz}),
$\Delta \Omega$ is the solid angle covered by a survey,
$d^2 V / d z d \Omega$ is the comoving volume element,
and $d n / d M$ is the cluster mass function (see Eq.~(\ref{eq:dndz})). 
We sum over redshift bins of size $\Delta z = 0.05$, between
$z_{\rm min}=0$ and $z_{\rm max} = 2.0$ for the DUO X--ray survey
and the SPT--like SZE survey, and between $z_{\rm min}=0.1$ and
$z_{\rm max} = 1.4$ for the LSST--like survey, although accurate
redshifts are not required for the $dN/dz$ test.

\begin{table}[htb]\small
\caption{Parameters for the Planck Survey.\label{planckparams}}
\begin{tabular}{|c|c|c|c|}
\hline
\hspace{5pt}Frequency (GHz)                             & 100  & 143  & 217  \\
\hline
\hspace{5pt}$\theta_c$ (arcmin)                         & 10.7 & 8.0  & 5.5  \\
\hspace{5pt}$\sigma_{c,T}$ ($\mu$K)                     & 5.4  & 6.0  & 13.1 \\
\hspace{5pt}$\sigma_{c,E}$ ($\mu$K)                     & $--$ & 11.4 & 26.7 \\
\hspace{5pt}$\ell_c$                                    & 757  & 1012 & 1472 \\
\hline
\end{tabular}
\end{table}

We construct the Fisher matrix for the redshift--space power
spectrum as~\cite{huha03}
\be
F_{\mu \nu}^{{\rm power}} = \sum_{i,j} \frac{\partial \ln (k_{\perp}^2 
k_{\parallel} P_c)_{ij}}{\partial p_{\mu}}
\frac{\partial \ln (k_{\perp}^2 k_{\parallel} P_c)_{ij}}
{\partial p_{\nu}} \frac{(V_k V_{\rm eff})_{ij}}{2},
\label{eq:powerfm}
\ee
where $P_c(k)$ is the cluster power spectrum. The two--dimensional $k$--space cells 
and the set of redshift bins are labeled by $i$ and $j$, respectively.
The factor of $(V_k V_{\rm eff} / 2)^{-1}$ estimates the uncertainty
$(\Delta P_c/P_c)^2 $ in the measured power spectrum, including the effects
of shot noise and cosmic variance~\cite{feldman94}. Here
$V_{\rm eff}$ is the effective volume probed by the survey
\be V_{\rm eff}(k) = \int \,d V_s \left[\frac{\bar{n}(z_j) P_c(k)}
{1 + \bar{n}(z_j) P_c(k)}\right]^2\,, \ee
where $\bar{n}$ is the expected average number density, 
and $V_k$ is the cylindrical volume factor in $k$--space:
\be V_k = \frac{2 \pi \Delta (k_{\perp}^2) \Delta k_{\parallel}}
{(2 \pi)^3}\,. \ee
We sum over $29^2$ linearly spaced $k$ cells from
$k_{\perp, \parallel} = 0.005$ to $0.15$ Mpc$^{-1}$, thus defining a cylinder
in three--dimensional $k$--space. We sum over redshift bins of size
$\Delta z = 0.2$, between $z_{\rm min}=0$ and $z_{\rm max} = 2.0$ for
the DUO--like survey and the SPT--like survey, and between $z_{\rm min}=0.1$ and
$z_{\rm max} = 1.4$ for the LSST--like survey. The choice of this relatively
wide redshift bin size is dictated by the need to have a sufficient
number of clusters in each bin for an accurate determination of the
power spectrum ($N \gsim 1,000$), as well as a wide enough $\Delta z$
that includes radial modes with $k_{\perp,\parallel} \approx 0.005$
Mpc$^{-1}$.

Finally, in addition to the constraints from clusters considered here,
we construct the Fisher matrix that can be used to forecast
cosmological parameter errors from the temperature and polarization
anisotropy of the cosmic microwave background (CMB). We have in mind a
near--future survey such as Planck~\cite{planck} that will measure
temperature and E--mode polarization auto--correlation (respectively,
TT and EE), as well as temperature--polarization cross--correlation
(TE). We neglect B--mode polarization. The full CMB Fisher matrix is
then given by~\cite{zaldarriaga96,zaldarriaga97}
\be F^{{\rm cmb}}_{\mu \nu} = \sum_\ell \sum_{X,Y}\frac{\partial C_{X,\ell}}{\partial p_{\mu}} 
 {\rm Cov^{-1}}(C_{X,\ell},C_{Y,\ell})\frac{\partial C_{Y,\ell}}{\partial p_{\nu}}\,, \ee
where $X,Y$ run over $TT$, $EE$ and $TE$ correlations. The covariance
matrix, ${\rm Cov}(C_{X,\ell},C_{Y,\ell})$, has elements
\begin{eqnarray}
\nonumber
 {\rm Cov}(C_{TT,\ell},C_{TT,\ell}) &=& \frac{2}{(2\ell+1)f_{\rm sky}}(C_{TT,\ell}+B_{T,\ell}^{-2})^2 \\
\nonumber
{\rm Cov}(C_{EE,\ell},C_{EE,\ell}) &=& \frac{2}{(2\ell+1)f_{\rm sky}}(C_{EE,\ell}+B_{E,\ell}^{-2})^2 \\
\nonumber
{\rm Cov}(C_{TE,\ell},C_{TE,\ell}) &=& \frac{1}{(2\ell+1)f_{\rm sky}}[C_{TE,\ell}^2 + \\
\nonumber
&& (C_{TT,\ell}+B_{T,\ell}^{-2})(C_{EE,\ell}+B_{E,\ell}^{-2})] \\
\nonumber
{\rm Cov}(C_{EE,\ell},C_{TE,\ell}) &=& \frac{2}{(2\ell+1)f_{\rm sky}}C_{TE,\ell}(C_{EE,\ell}+B_{E,\ell}^{-2}) \\
\nonumber
{\rm Cov}(C_{TT,\ell},C_{TE,\ell}) &=& \frac{2}{(2\ell+1)f_{\rm sky}}C_{TE,\ell}(C_{TT,\ell}+B_{T,\ell}^{-2}) \\
 {\rm Cov}(C_{TT,\ell},C_{EE,\ell}) &=& \frac{2}{(2\ell+1)f_{\rm sky}}C_{TE,\ell}^2 \,,
\end{eqnarray}
where $f_{\rm sky}$ is the fraction of the sky covered. The
$B_{T,\ell}$'s and $B_{E,\ell}$'s account for experimental noise for
temperature and polarization measurements, respectively, and are given
by~\cite{rocha03}
\begin{equation}
B_{\ell}^2 = \sum_c (\sigma_c\theta_c)^{-2} e^{-\ell(\ell+1)/\ell_c^2}\,,
\label{eq:channels}
\end{equation}
where the sum is over the different frequency channels denoted by $c$,
$\sigma_c$ is the sensitivity, $\theta_c$ is the beam width, and
$l_c\equiv 2\sqrt{2\ln 2}/\theta_c$ is the corresponding ``cut--off''
multipole. Equation \ref{eq:channels} assumes that different
channels provide independent constraints.  We follow
previous theoretical ``error forecast'' work in adopting this
assumption; however, we note that this implicitly assumes that all
foregrounds were perfectly removed from the temperature and
polarization maps.  In reality, imperfect removal of foregrounds will
induce correlations among the channels, which will have to be taken
into account in a refined analysis.  Modeling foregrounds and the
expected precision with which they can be removed is beyond the scope
of the present paper.

In this paper we focus on the Planck survey for concreteness. This
survey will measure temperature and polarization anisotropy in three
frequency bands, namely 100, 143 and 217~GHz, with fractional sky
coverage of $f_{\rm sky} \approx 0.8$. The parameters for this
experiment are listed in Table~\ref{planckparams} (taken
from~\cite{rocha03}). The various $C_\ell$'s are calculated up
to $\ell_{\rm max}=2,000$, as appropriate for Planck, using
KINKFAST~\cite{corasaniti04}, a modified version of
CMBFAST~\cite{seljak96} tailored for time-varying $w$.

\subsection{Fiducial Cosmology}

The Fisher matrix formalism estimates how well a survey can
distinguish a fiducial model of the universe from other models. The
results depend on the fiducial model itself. Throughout the paper, we
take a $7$--dimensional parameterization of a spatially--flat
($\Omega_k=0$), cold dark matter (CDM) cosmological model, dominated
by a cosmological constant ($\Lambda$).  The sensitivity of our
results to the choice of the fiducial parameters is discussed in
\S~\ref{sec:discussion} below.  The parameters are adopted from recent
measurements by the {\it Wilkinson Microwave Anisotropy Probe (WMAP)},
as summarized in Table~1 of~\cite{spergel03}: baryon density $\Omega_b
h^2 = 0.024$, matter density $\Omega_m h^2 = 0.14$, dark energy
density in units of the critical density $\Omega_{\rm DE} = 0.73$ (or
Hubble constant $H_0 = 100 h$ ${\rm km~s^{-1}~Mpc^{-1}}$ with $h =
0.72$), with present--day normalization $\sigma_8 = 0.9$ and scalar
power--law slope $n_s = 1$ of the primordial power spectrum. Following
Linder~\cite{linder03}, we parameterize the evolving dark energy
equation of state as
\be
w(z) = w_0 + w_a (1 - a) = w_0 + w_a \frac{z}{1 + z}\,, 
\label{wz}
\ee
with values in our fiducial model chosen to be $w_0 = -1$ and $w_a =
0$.  An alternative parameterization sometimes used in the literature
is $w(z) = w_0 + w_z z$. The errors we obtain here on $w_a$ should be
divided by approximately a factor of two to obtain the corresponding
errors on $w_z$. This follows from Taylor--expanding Eq.~(\ref{wz})
about $z = 1/2$, which is approximately where the sensitivity of
cluster surveys peak.

\subsection{Survey Parameters}
\label{subsec:param}

To determine the detection mass limit $M_{\rm min}(z)$ in
Eq.~(\ref{eq:dndz}), a mass--observable relation is needed.  We
consider three surveys for our analysis, a flux limited X--ray survey,
such as DUO, an SZE survey that is similar to the one to be carried
out with the South Pole Telescope, and a weak lensing survey that is
similar to that planned for the LSST. For the X--ray and SZE surveys,
we impose a minimum mass of $10^{14} h^{-1} M_{\odot}$ (if $M_{{\rm
min}}(z)$ as computed below turns out to be less than $10^{14} h^{-1}
M_{\odot}$) since less massive halos correspond to small clusters or
groups and are likely to depart from the scaling relations adopted
here.  We do not impose this lower bound for the weak lensing survey
since, in principle, the dark matter halos of low--mass clusters or
groups should still produce shear signals with a well--defined
mass--shear relation.

For the {\it DUO--like X--ray survey}, we adopt a bolometric
flux--mass relation of the form:
\be
f_x(z) 4 \pi d_L^2 = A_x M_{200}^{\beta_x} E^2(z) (1+z)^{\gamma_x}, 
\label{fx}
\ee
where $f_x$ in units of ${\rm erg~s^{-1}~cm^{-2}}$ is the bolometric
flux limit, $d_L$ in units of Mpc is the luminosity
distance, $M_{200}$ is the mass of the cluster, and
$H(z) = H_0 E(z)$ is the Hubble parameter at redshift $z$. Following
Majumdar \& Mohr~\cite{majumdar04}, we adopt $\log_{10}(A_x) = -4.159$,
$\beta_x = 1.807$, and $\gamma_x = 0$ as fiducial values. We model the DUO observations
as a combination of a ``wide'' survey, covering a sky area of
$\Delta \Omega = 6,000$ deg$^2$ with a bolometric flux limit of
$f_x > 1.75 \times 10^{-14}$~${\rm erg~s^{-1}~cm^{-2}}$ (corresponding
to $f_x > 7 \times 10^{-14}$~${\rm erg~s^{-1}~cm^{-2}}$ in the
$0.5:10$~keV band); and a ``deep'' survey, spanning
$\Delta \Omega = 150$ deg$^2$ with a bolometric flux limit of
$f_x > 2.25 \times 10^{-14}$~${\rm erg~s^{-1}~cm^{-2}}$
($f_x > 9 \times 10^{-15}$~${\rm erg~s^{-1}~cm^{-2}}$ in the
$0.5:10$~keV band). With these parameters, for our fiducial cosmological
model, the wide survey yields $\sim 10,000$ clusters while the deep
survey yields $\sim 1,500$ clusters. These numbers are consistent with
the existing data on the $\log N - \log S$ relation for clusters in
soft X--ray bands~\cite{gioia01} and also with independent estimates
for the total number of clusters expected to be detected by
DUO~\cite{majumdar04}.

For an {\it SPT--like SZE survey}, we adopt an SZE flux--mass relation:
\be
f_{sz}(z) d_A^2 = f(\nu) f_{\rm ICM} A_{sz} M_{200}^{\beta_{sz}}
E^{2/3}(z) (1+z)^{\gamma_{sz}},
\label{fsz}
\ee
where $f_{sz}$ in units of ${\rm mJy}$ is the observed flux decrement,
$d_A$ in units of Mpc is the angular diameter distance, $f(\nu)$ is
the frequency dependence of the SZE distortion, and $f_{\rm ICM}$ is
the mass fraction of the intra--cluster medium. We model the SPT--like
SZE survey as a flux--limited survey with $f_{sz} > 5$~mJy at $150$~GHz.  
While this is an oversimplification, the threshold value
approximately represents the total flux decrement of the smallest
cluster that can be detected at 5$\sigma$ significance with SPT.
\footnote{It has been pointed out~\cite{battye03} that clusters
are extended sources, and a convolution with SPT's beam profile leads
to a loss of sensitivity. This reduces in~\cite{battye03} the number
of clusters detectable with SPT to $\sim 4500$. However, this
reduction stems from the {\it point source} sensitivity of 5 mJy
adopted in~\cite{battye03}.  The threshold of 5 mJy we chose
represents the {\it total flux} of an extended cluster, and it can be
roughly understood as follows. The sensitivity of SPT at 150 GHz
within its 1 square arcminute beam is $\sim 10$~$\mu$K \cite{spt},
corresponding to 0.3 mJy.  The $\sim 2$~arcmin inner region of the
smallest detectable cluster fills $\sim 12$~beams, yielding a total
noise of $\sim 1$~mJy within the cluster aperture.}  We also adopt the
fiducial parameters $\log_{10}(A_{sz}) = 8.9$, $\beta_{sz} = 1.68$,
$\gamma_{sz}=0$, and covering $\Delta \Omega = 4,000~{\rm deg}^2$. We
also assume $f_{\rm ICM}=0.12$.  With these parameters, for our
fiducial cosmological model, this survey yields $\sim 20,000$
clusters.

Finally, for the {\it LSST--like WL survey}, we
follow Hamana et al. ~\cite{hamana03} to find a relation between the
dimensionless shear and halo mass, given by
\be 
\kappa_G = 
\alpha(\theta_G)
\left[\frac{M_{\rm vir}/(\pi r_s^2)}{\Sigma_{cr}}\right].
\label{sig-to-noi}
\ee
Here $\kappa_G$ is the average shear within a Gaussian filter of
angular size $\theta_G$; $M_{\rm vir}$ and $r_s$ are the mass and
scale radius of a cluster with an NFW density profile ($r_s = r_{\rm
vir} / \cnfw)$; $z_l$ is the redshift of the cluster; and $d_A$ is the
angular diameter distance to the cluster. The coefficient $\alpha$ is
given by
\be
\alpha = \frac {\int_0^\infty dx ~(x/x_G^2)
~\exp\left(-x^2/x_G^2\right) f(x)} {\log(1+\cnfw)-\cnfw/(1+\cnfw)},
\label{eq:alpha}
\ee
where $x = \phi / \theta_s$ a dimensionless angular coordinate, and
$x_G \equiv \theta_G / \theta_s$ corresponds to the smoothing scale,
with $\theta_s = r_s / d_A(z_l)$ denoting the angular scale radius.
The dimensionless surface density profile $f(x)$ is given by
equation~(7) in Hamana {\it et al.}~\cite{hamana03}. 
Finally, the mean inverse critical surface mass density is given by
\be
\Sigma_{{\rm cr}}^{-1} = \frac{4\pi G}{c^2} a(z_l) \chi(z_l)
\frac{\int_{z_l}^\infty dz ~{dn/dz} ~(1-\chi(z_l)/\chi(z))}
{\int_0^\infty dz ~{dn/dz}},
\ee
where $a$ is the scale factor and $\chi$ denotes the comoving radial
distance (valid for the flat universe with $\Omega_k=0$ we are
assuming).

We model the LSST-like survey to have a constant detection threshold
of $\kappa_G=4.5\sigma_{\rm noise}$~\cite{hennawi04}.  The noise is
given by the ratio of the mean ellipticity dispersion
($\sigma_\epsilon$) of galaxies and the total number of background
galaxies contained within a smoothing aperture
$\theta_G$~\cite{waerbeke00}, $\sigma_{\rm noise}^2 =
\sigma_\epsilon^2/(4\pi\theta_G^2 n_g)$. We adopt
$\sigma_\epsilon=0.3$ and a number density of background galaxies
$n_g=65$ arcmin$^{-2}$~\cite{song03}, and the
angular smoothing scale $\theta_G=1$ arcmin, which corresponds to optimal
S/N for the range of cluster masses and redshifts we considered.  We take the survey to cover
$\Delta \Omega = 18,000$ deg$^2$ and to extend over the cluster
redshift range $0.1 \leq z \leq 1.4$. This yields $\sim 200,000$
clusters for our fiducial cosmology.

\begin{figure}
\epsfxsize=3.3 in \centerline{\epsfbox{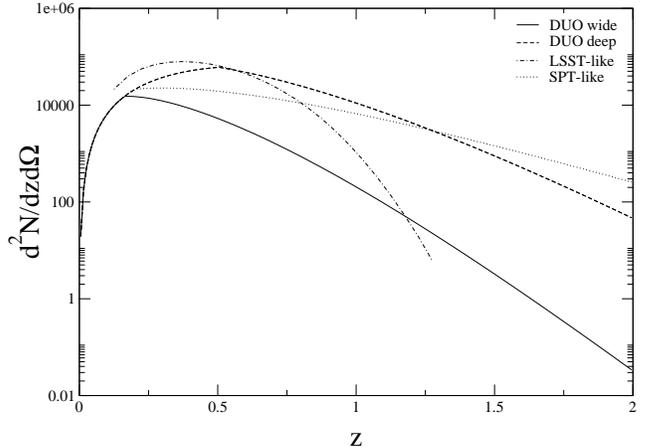}}
\caption{Expected number of clusters per redshift per unit solid angle for the fiducial cosmology.}
\label{fig:dndzcompare}
\end{figure}

Evidently, the various surveys considered here span different
redshift range, sky coverage and sensitivity. It is therefore useful
to compare the expected number of clusters per redshift and unit
solid angle for each survey for our fiducial cosmology. This is shown in
Fig.~\ref{fig:dndzcompare}. We see that the SPT--like survey is the most
sensitive probe at high redshift, a consequence of the fact that the
mass limit for SZE surveys is nearly redshift--independent. In comparison,
the DUO X--ray and LSST--like surveys drop more sharply with redshift.
Note that the fixed ``mass floor'' of $10^{14} h^{-1} M_{\odot}$
determines the number of clusters at $z<0.2, 0.25, 0.5$ for DUO wide,
SPT--like and DUO deep, respectively. In comparison, the LSST--like
counts are dominated by small clusters or groups with $M \leq 10^{14} h^{-1} M_{\odot}$.

\begin{table}[tb]\small
\caption{Estimated Cosmological Parameter Constraints from Planck.\label{cmb}}
\vspace{\baselineskip}
\begin{tabular}{|c|c|}
\hline
\hspace{5pt}{\bf Planck Survey}                         &        \\
\hspace{5pt}$\Delta\Omega_{\rm DE}$                     & 0.035  \\
\hspace{5pt}$\Delta\Omega_m h^2$                        & 0.0012 \\
\hspace{5pt}$\Delta\sigma_8$                            & 0.041  \\
\hspace{5pt}$\Delta w_0$                                & 0.32   \\
\hspace{5pt}$\Delta w_a$                                & 1.0    \\
\hspace{5pt}$\Delta\Omega_b h^2$                        & 0.00014\\
\hspace{5pt}$\Delta n_s$                                & 0.0035 \\
\hline
\end{tabular}
\end{table}

\begin{table}[tb]\small
\caption{Estimated Cosmological Parameter Constraints from DUO.  The $dN/dz$ column
includes priors from WMAP: $\Delta\Omega_b h^2 = 0.0010$, and $\Delta n_s = 0.040$.  \label{clustersonly}}
\begin{tabular}{|c|c|c|}
Survey and Parameter Constraints & \multicolumn{1}{|c|}{$dN/dz$} & \multicolumn{1}{|c|}{$P_c(k)$} \\
\hline
\hspace{5pt}{\bf DUO Wide (6,000 deg$^2$)}              &       &        \\
\hspace{5pt}$\Delta\Omega_{\rm DE}$                     & 0.14   & 0.037 \\
\hspace{5pt}$\Delta\Omega_m h^2$                        & 0.25   & 0.096 \\
\hspace{5pt}$\Delta\sigma_8$                            & 0.16   & 0.10  \\
\hspace{5pt}$\Delta w_0$                                & 0.16   & 0.59  \\
\hspace{5pt}$\Delta w_a$                                & 0.92   & 3.2   \\
\hspace{5pt}$\Delta\Omega_b h^2 $                       & 0.0010 & 0.023 \\
\hspace{5pt}$\Delta n_s$                                & 0.040  & 0.18  \\
\hline
\hspace{5pt}{\bf DUO Deep (150 deg$^2$)}                &        &        \\
\hspace{5pt}$\Delta\Omega_{\rm DE}$                     & 0.097  & 0.11  \\
\hspace{5pt}$\Delta\Omega_m h^2$                        & 0.33   & 0.25  \\
\hspace{5pt}$\Delta\sigma_8$                            & 0.040  & 0.25  \\
\hspace{5pt}$\Delta w_0$                                & 0.29   & 0.78  \\
\hspace{5pt}$\Delta w_a$                                & 2.5    & 3.7   \\
\hspace{5pt}$\Delta\Omega_b h^2 $                       & 0.0010 & 0.059 \\
\hspace{5pt}$\Delta n_s$                                & 0.040  & 0.49  \\
\hline
\end{tabular}
\end{table}

\begin{table*}[htb]\small
\caption{Estimated Cosmological Parameter Constraints from Clusters and CMB Combined.  The $dN/dz$ column
includes priors from WMAP: $\Delta\Omega_b h^2 = 0.0010$, and $\Delta n_s = 0.040$. \label{clustersandcmb}}
\hbox to \hsize{\hfil\begin{tabular}{|c|c|c|c|c|}
Survey and Parameter Constraints                & \multicolumn{1}{|c|}{$dN/dz$} & \multicolumn{1}{|c|}{$P_c(k)$} & \multicolumn{1}{|c|}{$dN/dz$ + $P_c(k)$} & \multicolumn{1}{|c|}{$dN/dz$ + $P_c(k)$ + Planck} \\
\hline
\hspace{5pt}{\bf DUO Combined}                          &       &        &        &             \\
\hspace{5pt}$\Delta\Omega_{\rm DE}$                     & 0.011  & 0.032  & 0.0074 & 0.0064  \\
\hspace{5pt}$\Delta\Omega_m h^2$                        & 0.022  & 0.084  & 0.0098 & 0.00041 \\
\hspace{5pt}$\Delta\sigma_8$                            & 0.016  & 0.088  & 0.012  & 0.011   \\
\hspace{5pt}$\Delta w_0$                                & 0.10   & 0.45   & 0.096  & 0.061   \\
\hspace{5pt}$\Delta w_a$                                & 0.48   & 2.3    & 0.45   & 0.19    \\
\hspace{5pt}$\Delta\Omega_b h^2 $                       & 0.0010 & 0.021  & 0.0010 & 0.00011 \\
\hspace{5pt}$\Delta n_s$                                & 0.040  & 0.15   & 0.033  & 0.0024  \\
\hline
\hspace{5pt}{\bf SPT-like Survey (4,000 deg$^2$) }                      &       &        &        &             \\
\hspace{5pt}$\Delta\Omega_{\rm DE}$                     & 0.036  & 0.033 & 0.014   & 0.0097  \\
\hspace{5pt}$\Delta\Omega_m h^2$                        & 0.049  & 0.056 & 0.0083  & 0.00027 \\
\hspace{5pt}$\Delta\sigma_8$                            & 0.031  & 0.064 & 0.018   & 0.012   \\
\hspace{5pt}$\Delta w_0$                                & 0.22   & 0.41  & 0.15    & 0.082   \\
\hspace{5pt}$\Delta w_a$                                & 0.59   & 1.8   & 0.46    & 0.18    \\
\hspace{5pt}$\Delta\Omega_b h^2 $                       & 0.0010 & 0.014 & 0.00099 & 0.00011 \\
\hspace{5pt}$\Delta n_s$                                & 0.040  & 0.094 & 0.029   & 0.0023  \\
\hline
\hspace{5pt}{\bf LSST-like Survey (18,000 deg$^2$)}                     &        &        &        &            \\
\hspace{5pt}$\Delta\Omega_{\rm DE}$                     & 0.0053 & 0.0080 & 0.0024  & 0.0023  \\
\hspace{5pt}$\Delta\Omega_m h^2$                        & 0.026  & 0.021  & 0.0048  & 0.00024 \\
\hspace{5pt}$\Delta\sigma_8$                            & 0.0035 & 0.022  & 0.0025  & 0.0024  \\
\hspace{5pt}$\Delta w_0$                                & 0.051  & 0.10   & 0.024   & 0.023   \\
\hspace{5pt}$\Delta w_a$                                & 0.086  & 0.47   & 0.077   & 0.061   \\
\hspace{5pt}$\Delta\Omega_b h^2 $                       & 0.0010 & 0.0050 & 0.00097 & 0.00010 \\
\hspace{5pt}$\Delta n_s$                                & 0.040  & 0.040  & 0.015   & 0.0022  \\
\hline
\end{tabular}\hfil}
\end{table*}

\begin{table*}[htb]\small
\caption{Parameter Constraints including Self--Calibration.  The $dN/dz$ column
includes priors from WMAP: $\Delta\Omega_b h^2 = 0.0010$, and $\Delta n_s = 0.040$.\label{clustersandselfcal}}
\hbox to \hsize{\hfil\begin{tabular}{|c|c|c|c|c|}
Survey and Parameter Constraints                & \multicolumn{1}{|c|}{$dN/dz$} & \multicolumn{1}{|c|}{$P_c(k)$} & \multicolumn{1}{|c|}{$dN/dz$ + $P_c(k)$} & \multicolumn{1}{|c|}{$dN/dz$ + $P_c(k)$ + Planck} \\
\hline
\hspace{5pt}{\bf DUO Combined}                          &       &        &        &             \\
\hspace{5pt}$\Delta\Omega_{\rm DE}$                     & 0.030  & 0.043  & 0.015  & 0.012   \\
\hspace{5pt}$\Delta\Omega_m h^2$                        & 0.14   & 0.091  & 0.0098 & 0.00067 \\
\hspace{5pt}$\Delta\sigma_8$                            & 0.058  & 0.12   & 0.016  & 0.013   \\
\hspace{5pt}$\Delta w_0$                                & 0.44   & 0.53   & 0.20   & 0.15    \\
\hspace{5pt}$\Delta w_a$                                & 1.2    & 2.5    & 0.82   & 0.46    \\
\hspace{5pt}$\Delta\Omega_b h^2 $                       & 0.0010 & 0.022  & 0.0010 & 0.00011 \\
\hspace{5pt}$\Delta n_s$                                & 0.040  & 0.18   & 0.034  & 0.0027  \\
\hspace{5pt}$\Delta \log A_x$                           & 0.27   & 0.18   & 0.050  & 0.037   \\
\hspace{5pt}$\Delta b_x$                                & 0.29   & 0.18   & 0.030  & 0.027   \\
\hspace{5pt}$\Delta \gamma_x$                           & 0.67   & 0.63   & 0.17   & 0.081   \\
\hline
\hspace{5pt}{\bf SPT-like survey}                       &       &        &        &             \\
\hspace{5pt}$\Delta\Omega_{\rm DE}$                     & 0.13   & 0.037  & 0.017  & 0.014   \\
\hspace{5pt}$\Delta\Omega_m h^2$                        & 0.65   & 0.077  & 0.0086 & 0.00063 \\
\hspace{5pt}$\Delta\sigma_8$                            & 0.14   & 0.099  & 0.019  & 0.017   \\
\hspace{5pt}$\Delta w_0$                                & 0.42   & 0.46   & 0.19   & 0.14    \\
\hspace{5pt}$\Delta w_a$                                & 2.4    & 1.8    & 0.81   & 0.40    \\
\hspace{5pt}$\Delta\Omega_b h^2 $                       & 0.0010 & 0.018  & 0.0010 & 0.00011 \\
\hspace{5pt}$\Delta n_s$                                & 0.040  & 0.15   & 0.032  & 0.0026  \\
\hspace{5pt}$\Delta \log A_{sz}$                        & 0.59   & 0.35   & 0.12   & 0.056   \\
\hspace{5pt}$\Delta b_{sz}$                             & 0.79   & 0.35   & 0.12   & 0.063   \\
\hspace{5pt}$\Delta \gamma_{sz}$                        & 1.7    & 0.62   & 0.19   & 0.057   \\
\hline
\end{tabular}\hfil}
\end{table*}

\begin{table*}[htb]\small
\caption{Calibrated Cosmological Parameter Constraints from LSST and CMB Combined.  The $dN/dz$ column
includes priors from WMAP: $\Delta\Omega_b h^2 = 0.0010$, and $\Delta n_s = 0.040$. \label{LSSTcmb}}
\hbox to \hsize{\hfil\begin{tabular}{|c|c|c|c|c|}
Parameter Constraints                & \multicolumn{1}{|c|}{$dN/dz$} & \multicolumn{1}{|c|}{$P_c(k)$} & \multicolumn{1}{|c|}{$dN/dz$ + $P_c(k)$} & \multicolumn{1}{|c|}{$dN/dz$ + $P_c(k)$ + Planck} \\
\hline
\hspace{5pt}{\bf LSST-like Survey}                      &        &        &        &            \\
\hspace{5pt}$\Delta\Omega_{\rm DE}$                     & 0.0081 & 0.012  & 0.0037  & 0.0033  \\
\hspace{5pt}$\Delta\Omega_m h^2$                        & 0.038  & 0.032  & 0.0059  & 0.00024 \\
\hspace{5pt}$\Delta\sigma_8$                            & 0.0054 & 0.034  & 0.0038  & 0.0037  \\
\hspace{5pt}$\Delta w_0$                                & 0.079  & 0.16   & 0.037   & 0.036   \\
\hspace{5pt}$\Delta w_a$                                & 0.13   & 0.72   & 0.12    & 0.093   \\
\hspace{5pt}$\Delta\Omega_b h^2 $                       & 0.0010 & 0.0077 & 0.00098 & 0.00010 \\
\hspace{5pt}$\Delta n_s$                                & 0.040  & 0.062  & 0.021   & 0.0022  \\
\hline
\end{tabular}\hfil}
\end{table*}

\section{Results}
\label{sec:results}

Before discussing our results for the cluster surveys, we first
summarize constraints from the CMB alone.  Our projections for the
Planck satellite, using the seven--parameter Fisher matrix, are listed in
Table~\ref{cmb} and are consistent with well--known previous
forecasts~\cite{zaldarriaga97,eisenstein98,hu02}.
This table shows the power of the CMB in constraining the matter and
baryon density, $\Omega_m h^2$ and $\Omega_b h^2$, respectively, as
well as the spectral tilt, $n_s$. However, as is well known, the
equation of state of the dark energy, parameterized by $w_0$ and
$w_a$, is poorly constrained by CMB observations alone~\cite{maor02,frieman03}.
This is because the dependence of the CMB power spectrum on these two
parameters comes mainly from the distance to last scattering,
$d_{\rm LS}$, which involves a double integral of $w(z)$:
\begin{equation}
\frac{d_{\rm LS}}{3000\;{\rm Mpc}} \approx \int_0^{z_{{\rm rec}}}\frac{dz}{\sqrt{\Omega_m h^2(1+z)^3 + (1-\Omega_m)h^2g(z)}}\,,
\label{dls}
\end{equation}
where $z_{{\rm rec}}$ is the redshift of recombination, and where
\begin{equation}
g(z)\equiv \exp\left\{3\int_0^{z}\frac{(1+w(z'))dz'}{1+z'}\right\}\,.
\end{equation}
Since $w_0$ and $w_a$ only appear inside this double integral, there
is a severe degeneracy that can leave $d_{\rm LS}$ nearly invariant under
changes in these parameters. 

Table~\ref{clustersonly} summarizes the results for the DUO wide and
deep surveys, both from cluster counts ($dN/dz$) and power spectrum
($P_c(k)$)~\footnote{In this Table, as well as in
Tables~\ref{clustersandcmb} and~\ref{clustersandselfcal} below, we
have included priors on $\Omega_b h^2$ and $n_s$ from the present
constraints on these parameters from WMAP (0.001 and 0.04,
respectively) for the $dN/dz$ Fisher matrix. The motivation of adding
these priors is that $dN/dz$ has essentially no sensitivity to these
two parameters, resulting in large predicted uncertainties which obscure
the information content of $dN/dz$.  The $dN/dz$ and $dN/dz+P(k)$ columns in all three
tables reflect these WMAP priors. In the last column, when CMB
information from Planck is added in Tables~\ref{clustersandcmb}
and~\ref{clustersandselfcal} below, the WMAP priors become
irrelevant.}.  This table addresses the issue of the relative merits of
survey size versus depth. Starting with the counts, we see that the
constraints for the wide and the deep surveys are of the same order,
even though the latter yields about 7 times fewer clusters. This is
because the deep survey, despite its limited angular coverage,
measures a higher fraction of high--redshift clusters. For the power
spectrum, however, the constraints are most sensitive to the total
number of clusters. Indeed, the errors on most parameters differ by
roughly a factor of $\sqrt{N_{\rm wide}/N_{\rm deep}}\approx 2.6$,
where $N_{\rm wide}$ and $N_{\rm deep}$ are the total number of
clusters for the respective surveys. The power spectrum $P_c(k)$
delivers good constraints on the densities ($\Omega$'s) and on
$\sigma_8$ in the wide survey, but has little sensitivity to $w_0$ and
$w_a$.
 
Table~\ref{clustersandcmb} addresses the issue of the relative merits
of $dN/dz$, $P_c(k)$, their combination, and their combination with
CMB anisotropy data from Planck, for all three surveys. The top third
of Table~\ref{clustersandcmb} lists the results of combining DUO wide
and deep. Of particular interest is the third column from the left,
which shows the constraints obtained by adding the Fisher matrices for
$dN/dz$ and $P_c(k)$ for the combined survey. These are, in short, the
most optimistic error bars from DUO alone.  The table also illustrates
the power of combining cluster counts with two--point function
statistics.  Indeed, the combined error bars (column 3) for
$\Omega_{\rm DE}$ and $\Omega_m h^2$ are about two times smaller than
those derived from either $dN/dz$ (column 1) or $P_c(k)$ (column 2)
alone. Finally, we see in the last column that combining DUO and
Planck further reduces the uncertainty on $w_0$ and $w_a$ by
approximately a factor of two. This underscores the complementarity of
cluster and CMB data in uncovering the nature of the dark energy.

The middle third of Table~\ref{clustersandcmb} shows our results for the
SPT--like SZE survey. Overall, the constraints on the cosmological
parameters are similar to those available from the DUO--like survey. 

The bottom third of Table~\ref{clustersandcmb} shows our estimated
parameter uncertainties for the LSST--like cluster survey. Comparing
the third column with the previous two again confirms the power of
combining counts with power spectrum. The constraints on $w_0$
and $w_a$ are of the order of a few percent. These remarkably tight
bounds (comparing favorably with those from the  Planck survey [Table~\ref{cmb}] for all
cosmological parameters except $\Omega_mh^2$, $\Omega_bh^2$ and $n_s$)
are the result of the very high cluster yield of this survey.  To
examine the sensitivity of these results to the inclusion of the
lowest mass clusters, we follow a more conservative approach by
imposing a minimum mass of $2 \times 10^{14} h^{-1} M_{\odot}$. This
reduces the number of clusters to $\sim 50,000$. The constraints on
$w(a)$ from LSST alone degrade by a factor of about two (consistent
with $\sqrt{N}$ scaling of statistical errors), to $\Delta w_0 =
0.030$ and $\Delta w_a = 0.23$. When combined with Planck, the errors
are nearly unaffected: $\Delta w_0 = 0.024$ and $\Delta w_a =
0.070$. Therefore we are confident that with enough clusters and
combining with Planck, LSST can constrain $w_a$ to a few percent.
Finally, we find that it is essential to include the cosmology--dependence of the
limiting mass for the LSST survey.  Repeating our analysis adopting the
redshift--dependent mass limit from the fiducial cosmology, and not
allowing it to vary with cosmology, results in an increase by a factor
of 3--4 in the uncertainties.

\begin{figure*}[htb]
\hbox to \hsize{\hfil \epsfxsize=6 in \centerline{\epsfbox{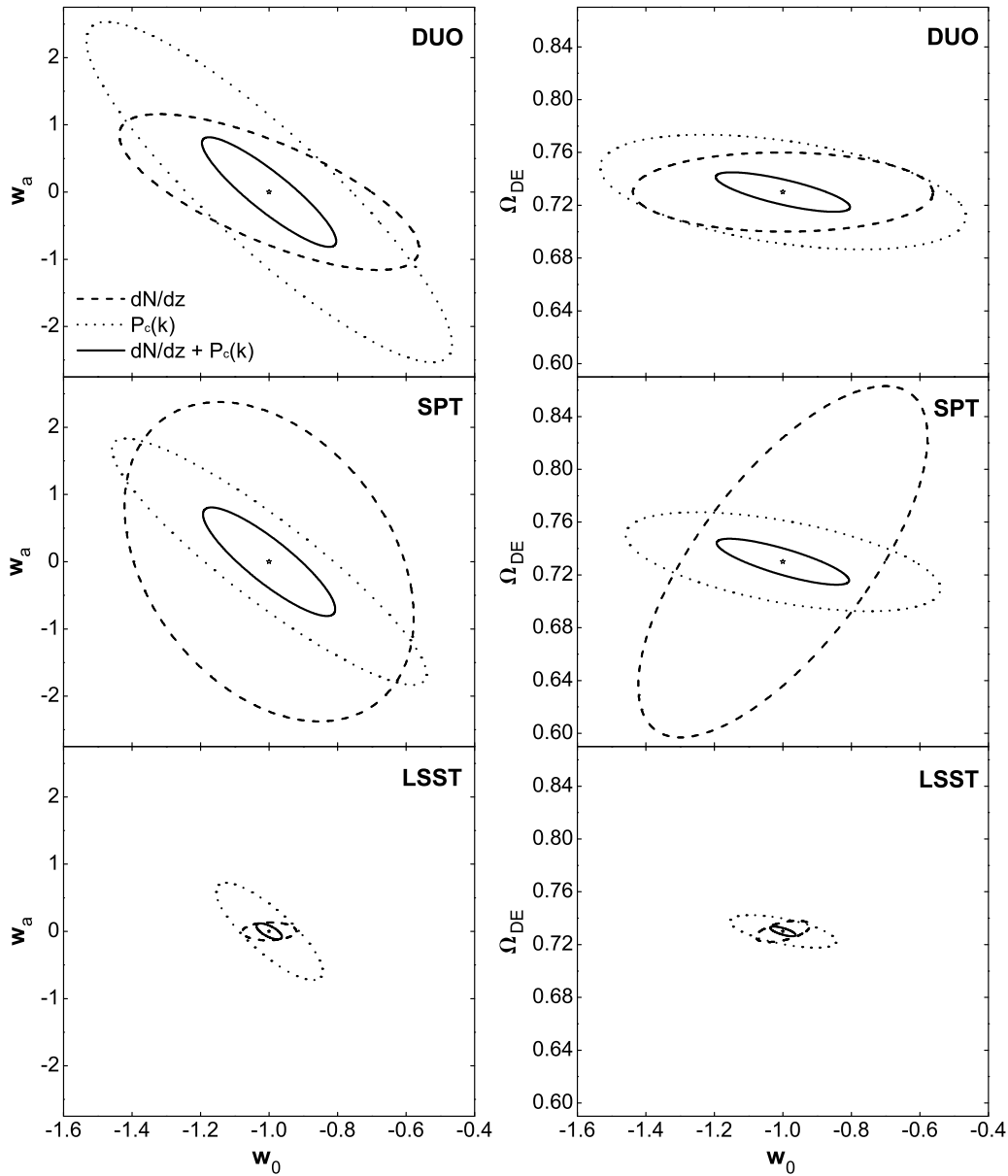}} \hfil}
\vspace{-3\baselineskip}
\caption{Constraints on dark energy parameters:
$w_0$ and $w_a$ (left), $w_0$ and $\Omega_{\rm DE}$ (right) for a DUO--like
X--ray survey (top), an SPT--like SZE survey (middle), and an LSST--like
weak lensing survey (bottom). The three curves in each figure show the
constraints available from $dN/dz$ (dashed), $P_c(k)$ (dotted),
and from their combination (solid). The star--shaped symbol at the center
of each figure indicates our fiducial cosmology.
The constraints for X--ray survey and
SZE survey are calculated for the self--calibration case. The constraints
are marginalized over all other cosmological and relevant structure parameters. In all cases, the
constraints from the combination of $dN/dz$ and $P_c(k)$ are at least a
factor of two stronger than from either method alone.}
\label{fig:w0wa}
\end{figure*}

\begin{figure*}[htb]
\hbox to \hsize{\hfil \epsfxsize=6 in \centerline{\epsfbox{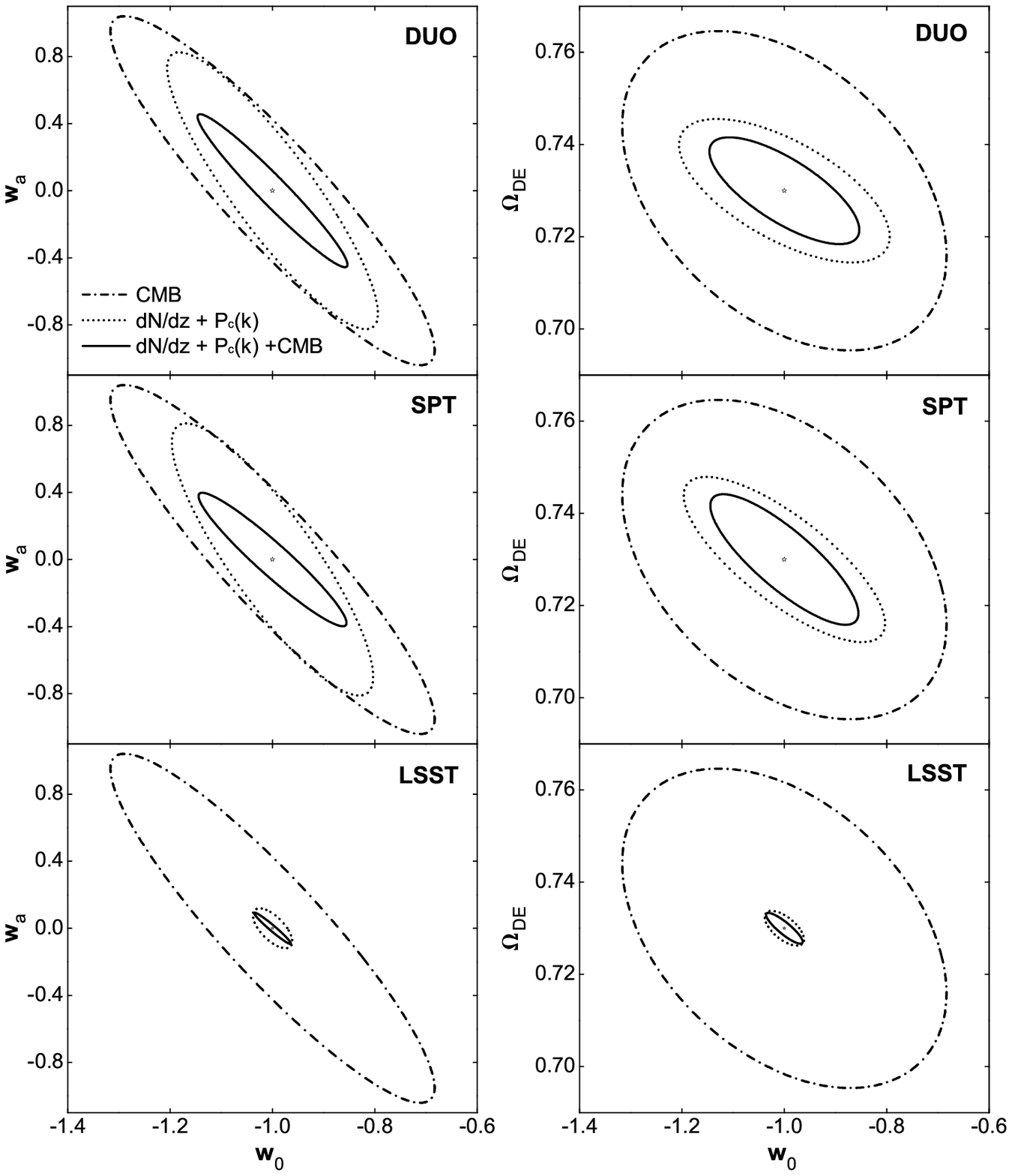}} \hfil}
\vspace{-3\baselineskip}
\caption{Constraints on dark energy parameters:
$w_0$ and $w_a$ (left), $w_0$ and $\Omega_{\rm DE}$ (right), by
combining a DUO--like X--ray survey (top), an SPT--like SZE survey (middle),
or an LSST--like weak lensing survey (bottom) with Planck--like CMB
observations. The three curves in each figure show the constraints
available from the combination of $dN/dz$ and $P_c(k)$ (dotted), CMB
alone (dot--dashed), and from the combination of all of the
above (solid). As in Fig.~\ref{fig:w0wa}, the constraints
for X--ray survey and SZE surveys are calculated for the 
self--calibration case, and constraints are marginalized
over all other relevant parameters. Note the different
scales of the horizontal and vertical axes compared to Fig.~\ref{fig:w0wa}.}
\label{fig:w0wacmb}
\end{figure*}

In Table~\ref{clustersandselfcal}, we repeat the analysis for the DUO
and SPT--like surveys, but this time taking into account the
uncertainty in the structure and evolution of clusters. In other
words, we require that the cluster surveys constrain not only the
cosmology, but also the parameters of the mass--observable
relation. For DUO, this is modeled by including the parameters $A_x$,
$\beta_x$ and $\gamma_x$ of Eq.~(\ref{fx}) in the Fisher matrix
analysis. Similarly, for the SPT--like survey, we include $A_{sz}$,
$\beta_{sz}$ and $\gamma_{sz}$ from Eq.~(\ref{fsz}). For both
surveys, we see that adding self--calibration increases the error on
$w_0$ and $w_a$ by a factor of $\lsim 2$ in comparison with the
corresponding results in Table~\ref{clustersandcmb}.  Overall, we see
that including self--calibration parameters still yields very good
constraints on the cosmology. The constraints from $dN/dz+P_c(k)$ and
$dN/dz+P_c(k)+\rm Planck$ are shown graphically in
Figs.~\ref{fig:w0wa} and~\ref{fig:w0wacmb}, including
self--calibration for the DUO--like and SPT--like surveys.

Coming back to Table~\ref{clustersandcmb} and comparing the first
column from the left ($dN/dz$) with the third ($dN/dz +
P_c(k)$), we note that, for DUO, adding the power spectrum does not
significantly tighten the constraints on $w_0$ and $w_a$ in this
case. However, for the ``self--calibration'' case
(Table~\ref{clustersandselfcal}), combining the two methods helps
greatly on all the constraints. For SPT--like and LSST--like surveys,
combining two methods always gives stronger constraints. This is
clearly illustrated in Fig.~\ref{fig:w0wa}.

An obvious method to cross-check for systematic effects due to cluster
structure and evolution, and to improve constraints, is to combine the
X-ray and SZE data.  We have found that the uncertainties in $w_0$ and
$w_a$ reduce to 0.083 and 0.35 when the self-calibrated SPT and DUO
samples are considered in combination, and to 0.060 and 0.22  when
Planck is added to this combined sample.

Uncertainties in cluster structure and evolution should be less
severe for WL signatures, which probe the dark matter potential
directly. However, to compare the WL error forecasts more fairly with
the SZ and X--ray predictions, in Table~\ref{LSSTcmb} we show LSST
predictions that incorporate two additional uncertainties inherent to
WL selection: false detections and incompleteness. We relied on
published results of numerical simulations~\cite{hamana03,hennawi04} to
calibrate our results: we assumed $25\%$ of detections are false, and
$30\%$ of real clusters are undetected. These effects increase the
parameter errors, relative to those in the bottom third of
Table~\ref{clustersandcmb} by $\approx 50 \%$ (except for $\Omega_b
h^2$ and $n_s$, which are significantly determined by the WMAP priors
and are less affected). See the next section for a more detailed
discussion.

\section{Discussion}
\label{sec:discussion}

In the previous section, we derived constraints on cosmological
parameters from future SZE, X--ray, and WL surveys.  Our Fisher matrix
approach should be interpreted as yielding lower limits on the
achievable statistical errors.  In our analysis, we adopted unique
relations between the observables and cluster mass, which come from
simple models of cluster structure and evolution. We required the SZ
and X--ray surveys to ``self--calibrate'' and constrain structure and
evolution parameters simultaneously with cosmology. However, the
power--law form of the relations we adopted will have to be further
constrained. This should be feasible by combining the three
observables in the different wave bands for a subset of the samples,
and by adding new observables (such as the shape of the mass function,
the angular size, velocity dispersion) which we have not considered
here.  In the case of the WL sample, we relied on results from
numerical simulations to calibrate the mass--observable relation, an
approach that can be refined with a larger suite of simulations in the
future.

The Fisher matrix technique, especially the way of combining independent
constraints by summing individual Fisher matrices, allows us to explore
the physical origin of the cosmological information. A glance at Eqs.~(\ref{eq:countfm})
and~(\ref{eq:powerfm}) reveals that cosmology enters through several
physical quantities into the Fisher matrices, such as the cosmic volume,
growth function, transfer function, bias, etc. Unfortunately, the
marginalization over all other parameters involves a nonlinear inversion
of the Fisher matrix, which makes isolating the various sources of information
difficult. As an example, we repeated the analysis above, but
keeping either the volume factor, $k_{\perp}^2 k_{\parallel}$, or the bias parameter 
at their values in the fiducial model ({\it i.e.}, excluding their
derivatives from computing power spectrum Fisher matrices). We find that
as a result, the constraint on some of the parameters improve,
while others degrade, which does not offer a useful description of the
amount of information the volume factor or the bias parameter provides.

In addition, we explored the implicit assumption made above that the
bias parameter $b$ can be precisely modeled. We take a similar
approach to that used in addressing ``self--calibration'', by modeling
the bias as $b = b_0(z) (1+z)^{\gamma_b}$.  We effectively include an
additional new parameter, $\gamma_b$, in the Fisher matrix
analysis. Note that a constant factor of normalization would be
degenerate with $\sigma_8$.  In the fiducial model, $b_0(z)$ follows
from Eq.~(\ref{eq:biasz}), and $\gamma_b = 0$. The constraints
from cluster power spectrum including the ``non--standard evolution
parameter'' $\gamma_b$ causes only a minor degradation (under $10\%$)
of the constraints on $w_0$ and $w_a$ (relative to column 2 in
Table~\ref{clustersandcmb}), suggesting that the bias factor did not
drive the cosmological constraints we derived from the power
spectrum.

The results from the power spectrum also depend on the assumption of
the extent of the linear regime. To quantify the importance of the
small--scale modes, for the example of the SPT--like survey, we
decreased $k_{\rm max}$ from $0.15$ to $0.10$ Mpc$^{-1}$.  We found
that this degrades the error bars by a factor of up to $\approx 1.4$.
Similarly, decreasing $k_{\rm max}$ further to $0.075$ Mpc$^{-1}$
degrades all the error bars by a factor of up to $\approx 2$.  (See
also \cite{huha03} for a detailed discussion of the dependence of the
constraints separately on $k_{\perp,\rm max}$ and $k_{\parallel,\rm
max}$.)

In Table~\ref{clustersandcmb}, we have assumed that the
mass--observable relations apply exactly.  However, for real surveys,
the mass--observable relation is likely to have a non--negligible
scatter. We estimated the magnitude of scatter that would bias the
inferred parameters by an amount comparable to their statistical
errors, for the example of the SPT survey~\cite{battye03}.  We assumed that at a fixed
flux $f_{sz}$, the mass $M_{180}$ has a Gaussian distribution, with a
mean given by Eq.~\ref{fsz} (but converted from $M_{200}$ to
$M_{180}$), and a fractional r.m.s. deviation of
$\delta M/M=\sigma_M$.  We then re--computed the cluster abundance
$dN/dz$ (which is increased in the presence of scatter). We found that
at $\sigma_M\approx 7\%$, the change caused in the total number of
detectable clusters is comparable to the $\sqrt{N}$ error that is used
to obtain the constraints on cosmological parameters. A scatter
larger than $\sim 7\%$ would therefore be important, and it would have
to be modeled (i.e. by parameterizing the scatter, calibrated directly
from observations) in the analysis of real data.

A further complication, likely most relevant for the WL survey,
is the presence of false detections.  In our analysis of the
LSST--like survey, we used a constant shear S/N to select clusters,
which directly measures how well the shear signal stands out above the
ellipticity shot noise of the background galaxies.  False detections
can result from statistical fluctuations in these ellipticities.
Furthermore, physical structures, projected along the line of sight to
a given cluster, constitute additional background noise, and can
result in a false detection of two or more mass concentrations
(projected along the line of sight but physically separated in
redshift). Hennawi \& Spergel~\cite{hennawi04} have done a
comprehensive study of mass--selected clusters using $N$--body
simulations (see also~\cite{hamana03,white02} for other numerical studies of WL
cluster selection).  Defining {\it efficiency} as the fraction of the
peaks in the mass map that corresponds to real clusters, and {\it
completeness} as the fraction of real clusters we can detect, they
find in their simulation that for a $4.5$ standard deviation detection
threshold, $\approx 75\%$ of the detected clusters are real and
$\approx 70\%$ of the clusters can be detected~\cite{hamana03}.  They
emphasize that the expected cluster distribution, including false
detections and missing clusters, can be reliably calculated for any cosmological model
since the simulations depend only on gravity.  Thus, false detections
can be subtracted, and their presence serves only to increase the
statistical error. These effects would increase the parameter errors in
the bottom third of Table~\ref{clustersandcmb} by $\approx 50\%$
(summing in quadrature the errors on the number of real clusters and
on the number of false--detections, $\sqrt{((1/e-1)+1/e)/c}-1 \approx 50\%$,
where $e = 75\%$ is the efficiency and $c = 70\%$ is the completeness as we defined above).  

We note that the study in~\cite{hennawi04} utilized only the
``tomographic redshifts'' derived from the shear map itself (using
photometric redshifts only for the background galaxies), which have
large uncertainties. However, it should be feasible to routinely
measure photometric redshifts of the cluster member galaxies, as well,
and this additional information could be used to reduce the rate of
false detections.

Obtaining redshifts is a general issue relevant for all three
future surveys. While accurate (spectroscopic) redshifts are not
required for either the $dN/dz$ or $P(k)$ tests, photometric redshift
estimates will be required to utilize the distribution of clusters in
$z$, and also to use high--$k_\parallel$ modes of the power spectrum.
In the case of the LSST survey, we adopted a maximum redshift $z_{\rm
max} = 1.4$, the redshift out to which it should be feasible to obtain
photometric redshifts with the planned filters and
sensitivities~\cite{lsst}.  In the case of DUO, most of the clusters
in the wide survey ($\gsim 80\%$) will have photometric redshifts from
the overlapping Sloan Digital Sky Survey (SDSS) galaxies; the redshifts of the more distant
galaxies in the deep survey will be obtained in optical follow--up
programs.  Obtaining photometric redshifts of a large fraction of
clusters in the SPT survey beyond $z\gsim 1.4$ will be challenging.
However, we have re--computed our results for SPT ignoring all
clusters beyond $z=1.4$, and found that none of the constraints
degraded by more than a few percent.

It has often been argued that weak lensing only involves
gravitational deflection of light, so the mass--shear relation can be
determined without fully understanding the complex baryonic
physics. Clearly this is an idealization, since baryons can cool and
contract to establish a more centrally--condensed density profile than
the dark matter, altering the total gravitational field. To quantify
the resulting change in the WL shear, we followed a simple spherical
model in ~\cite{white04}, in which a fixed fraction $f_{\rm cool}$ of
baryons inside a `cooling radius' are allowed to cool and condense by
a further factor of $\sim 10$ after virialization.  The effect of such
a ``cooling flow'' on the mass profiles of halos is relatively small,
except in the inner regions.  In models with $f_{\rm cool}$ up to 30
\%, we find that the change in the Gaussian--averaged shear $\kappa_G$
can be up to $15\%$. This reflects the enhanced effect of baryons in
the core of the cluster, which is emphasized by the Gaussian
weighting.  However, this bias only affects the initial selection of
clusters. In practice, one will go back and extract the total mass of
the cluster from the shear map in a separate
analysis~\cite{fahlman94}.  We find that the total mass enclosed
within the virial radius is changed by a much smaller amount,
$<1\%$.

In this work, the constraints on cosmological parameters were derived
assuming a fiducial model with a cosmological constant as the dark
energy, {\it i.e.}, $w_0=-1$ and $w_a=0$. As mentioned above, the
choice of the fiducial model can, in principle, strongly affect the
predicted errors. To assess the robustness of our derived errors, we
repeated the analysis for a fiducial model with a strongly
time--dependent equation of state, namely $w_0=-0.8$ and $w_a=0.3$,
which is also consistent with current data. We find in this case that
all error bars derived from $dN/dz$ and $P(k)$ remain essentially
unchanged, except for $\Delta w_a$, which shrinks by about
25\%. Therefore, by choosing a pure $\Lambda$ model as our fiducial
cosmology, we obtain a conservative estimate of the error on $w_a$.

\section{Conclusions}
\label{sec:conclusions}

Three main lessons can be drawn from our work. First, the above
results provide further support for the usefulness of clusters as an
independent and important probe of the dark energy. At a general
level, the main advantage of clusters over other measurements lies in
the fact that the physics involved is predominantly gravitational and
hence simple; moreover, the observations are at low redshift and
therefore straddle the epoch of cosmic acceleration.

We have seen that even cluster counts ($dN/dz$) alone can yield
comparable or even better constraints on $w_0$ and $w_a$ than a
full--sky, high--precision CMB survey such as Planck. Our constraints
our summarized in Table~\ref{clustersandcmb}.  Of course, this assumes
perfect knowledge of the mass--observable relation, {\it i.e.}, that
the clusters are standard candles. We have shown, however, that even
after allowing for uncertainty in cluster evolution and structure, and
using the survey itself as a calibration tool, the errors on $w_0$ and
$w_a$ still compare respectably well with Planck, as seen from the
third column of Table~\ref{clustersandselfcal}. The constraints from
cluster surveys also compare favorably with other cosmological
probes. For instance, current constraints from type Ia supernovae are
on the order of $\Delta w_0 \approx 0.3$ and $\Delta w_a \approx
1.6$~\cite{reiss04}, assuming a strong prior of $\Delta\Omega_m=0.04$
from other measurements. The expected errors from the {\it Supernovae
Acceleration Probe} (SNAP) are $\Delta w_0 \approx 0.08$ and $\Delta
w_a \approx 0.3$, with the same prior on
$\Omega_m$~\cite{SNAP}. Relaxing this prior results in errors on the
order of $\Delta w_0 \approx 0.2$ and $\Delta w_a \approx
1.0$~\cite{maor02}, comparable to our projected constraints from DUO
or SPT.

A second lesson we wish to draw is the power of combining different
cluster observables in constraining cosmological parameters, in this
case cluster counts ($dN/dz$) and two--point statistics ($P_c(k)$).
The most striking illustrations of this are the LSST--like survey
(see Table~\ref{clustersandcmb}) where $\Delta w_0$ is a factor of
$\sim 2$ smaller after combining the two methods; and the SPT--like
self--calibrated case (see Table~\ref{clustersandselfcal}) where both
$\Delta w_0$ and $\Delta w_a$ shrink by a factor of $\gsim 2$.  While
we have focused here on $dN/dz$ and $P_c(k)$, other cluster
observables that could be included are three--point correlation
functions, assuming a large enough survey, information from the shape
of the mass function, $dN/dM$, as described in~\cite{hu03}, or scaling
relations~\cite{vhs03}.  We leave a more comprehensive study of such
additional cluster observables for future work.

Finally, our work underscores the complementarity of cluster
observables to CMB anisotropy. In general, we have seen that
adding the information from a Planck--like survey reduces $\Delta w_0$
and $\Delta w_a$ by a factor of 2 or so (although the LSST--like WL
survey has already very small errors, and adding Planck data improves
the constraints by a more modest amount). Thus clusters not only
constitute a powerful cosmological probe in their own right, but also
help in alleviating CMB observations from some of their well--known
degeneracies.

Future work includes the addition of other cluster observables, as
mentioned earlier, as well as using cluster surveys to place useful
constraints on the neutrino mass~\cite{upcoming}.

\acknowledgments We thank Joseph Hennawi, Wayne Hu, Joseph Mohr, and
the members of the DUO team for insightful discussions, Jochen Weller
for useful comments on the manuscript, Gil Holder and Joseph Mohr
for useful comments on SPT's sensitivity, Subhabrata Majumdar for
assistance in cross--checking our numerical codes, Pier--Stefano
Corasaniti for allowing us to use his modified version of CMBFAST,
and an anonymous referee whose comments improved this
paper. This work is supported in part by the US Department of Energy
under Contract No. DE-AC02-98CH10886, by the Columbia University
Academic Quality Fund and the Ohrstrom Foundation (JK).

\end{document}